\documentclass[twocolumn,aps,prb,superscriptaddress,longbibliography]{revtex4-2}
\usepackage{graphicx}
\usepackage{color}
\usepackage{amsmath}
\usepackage[export]{adjustbox}
\usepackage{enumitem}
\usepackage{amssymb}
\usepackage[normalem]{ulem}
\usepackage{adjustbox}
\usepackage{hyperref}
\hypersetup{
	colorlinks = true,
	linkcolor = blue,
	citecolor = blue,
	urlcolor  = blue,
}

\newcommand{\be}{\begin{equation}}
	\newcommand{\ee}{\end{equation}}

\newcommand{\bea}{\begin{eqnarray}}
	\newcommand{\eea}{\end{eqnarray}}

\newcommand{\p}{\partial}

\renewcommand{\vec}[1]{{\boldsymbol #1}}
\renewcommand{\epsilon}{\varepsilon}

\def\nn{\nonumber\\}

\begin{document}

\title{Dynamical localization in 2D topological quantum random walks}

	\date{\today}
	
	\author{D. O. Oriekhov}
	\affiliation{Kavli Institute of Nanoscience, Delft University of Technology, 2628 CJ Delft, the Netherlands}
	\author{Guliuxin Jin}
	\affiliation{Kavli Institute of Nanoscience, Delft University of Technology, 2628 CJ Delft, the Netherlands}%
	\author{Eliska Greplova}%
	\affiliation{Kavli Institute of Nanoscience, Delft University of Technology, 2628 CJ Delft, the Netherlands}

	\begin{abstract}
We study the dynamical localization of discrete time evolution of topological split-step quantum random walk (QRW) on a single-site defect starting from a uniform distribution. 		
Using analytical and numerical calculations, we determine the high localization probability regions in the parameter space of the quantum walker. These regions contain two or more pairs of trapped states, forming near a lattice defect. By investigating the spectral properties of the discrete-time evolution operators, we show that these trapped states have large overlap with the initial uniformly distributed state, thus offering a simple interpretation of the localization effect.
As this localization scheme could be interpreted as a variation of spatial quantum search algorithm, we compare the localization probability and time with other types of two-dimensional quantum walks that do not have topological phases and realize localization time scaling similar to Grover's algorithm. Finally we show that mechanism of localization we identified is robust against the influence of disorder. 

	\end{abstract}
	\maketitle
\section{Introduction}
The quantum random walks were introduced in Ref.~\cite{Aharonov1993} as a natural quantum version of classical random walks. Due to interference effects and the state superposition, the average travel path of walkers is much longer than their classical counterpart. The clear structure of possible physical systems that could realize the quantum walks~\cite{Aharonov1993} attracted much attention. In particular, it was shown that quantum walks are a possible way to implement Grover's algorithm~\cite{Grover1996} in the case of absence of high-quality multi-qubit quantum computer~\cite{Shenvi2003PRA,Childs2004,Ambainis2004,Inui2004PRA,Reitzner2009,Tulsi2008PRA}. It was also shown that the so-called Grover's oracle operator has a natural realization in quantum walks and is related to dynamical localization phenomena~\cite{Shenvi2003PRA,Childs2004,Ambainis2004,Inui2004PRA,Reitzner2009,Tulsi2008PRA}. In particular, the discrete-time quantum random walks are defined as a set of shift and coin-flip operators applied to the whole system at each time step~\cite{Ambainis2004,Endo2021,Aharonov1993,Shenvi2003PRA}. The successful search event is defined as a time of maximum localization of quantum walker on a defect when starting from a uniform distribution over the whole system. Going forward, we will refer to the
the time of appearance of the first maximum in localization probability as search time.
 \begin{figure}[t]
		\centering
		\includegraphics[scale=0.7]{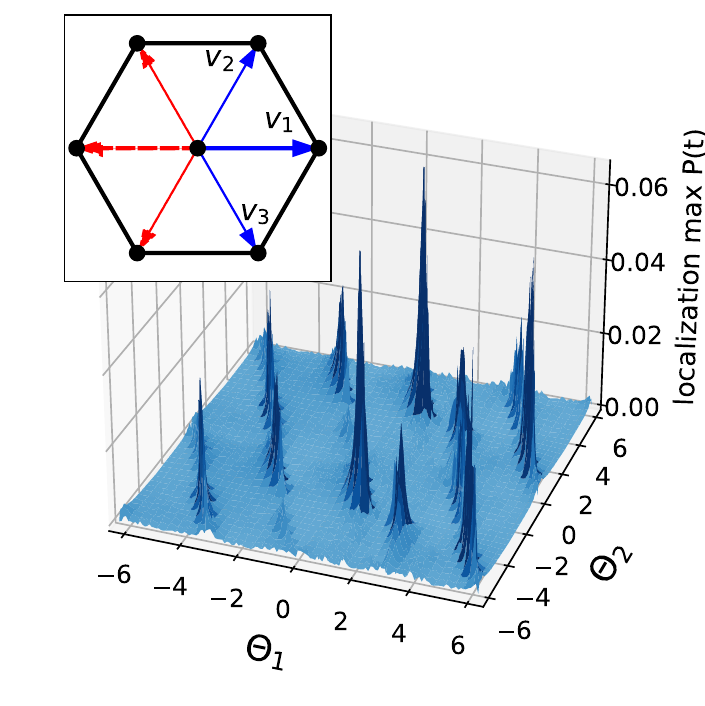}
		\caption{The localization probability of topological quantum walk as a function of spin rotation parameters $\Theta_{1,2}$ (see Eq.~\eqref{eq:QRW-definition}) for a defect with $\Theta_{1}^{def}=5\pi/8$, $\Theta_{2}^{def}=\pi/2$. High peaks are of two orders of magnitude larger than average localization probability and correspond to parameters for which quantum walker performs search efficiently. Inset: geometry of split-step QRW on a triangular lattice with three shift vectors $\vec{v}_{i}$ - solid ones correspond to shift of spin up components, dashed $-\vec{v}_{i}$  for spin down.}
		\label{fig1-max-map}
	\end{figure}

In the most common mechanism of localization, at least several eigenstates of QRW evolution operator create a constructive superposition on a defect at a certain time \cite{Inui2004PRA,Ambainis2004}. Little is known about the precise values of optimal parameters of defect that maximize localization relative to the surrounding system. In addition, it was shown that certain positioning of defects could significantly decrease localization probability \cite{Roget2023PRR}.
 Here, we study the two-dimensional split-step quantum random walks analogous to Chern type topological insulators in crystalline solids~\cite{Kitagawa2010PRA}. For such a quantum walk, it could be expected that topology creates a new mechanism of localization, similar to in-gap states in solids. Following this motivation we explore how topology affect the localization mechanism. The topological  split-step QRWs in 1D space were studied in Refs.`\cite{Endo2017,Endo2021}. Here, we analyze the parameter spaces of the quantum walker and the defect. We find that there exist sets of parameters that deterministically demonstrate a high localization probability at a given defect (see a localization probability map in Fig.~\ref{fig1-max-map}). In Sec.~\ref{sec:model} we show that these regions in parameter space can be categorized into two distinct classes based on their position inside the topological phase or on the phase separation lines. Notably, we find the search time for the parameters inside both classes of regions demonstrates the saturation to a constant with growing system size. The saturation of search time happens at different system sizes, while before this saturation the square root time growth behavior takes place.

The localization after constant time, regardless of system size, was previously identified only for electric Dirac QRW with non-local marking of defect in the form of Coulomb potential~\cite{Zylberman2021Entropy,Fredon2022Entropy}. The non-locality of defects creates an important difference from our scenario: our protocol does not require a large number of nodes to form a defect structure. In the algorithmic terminology, in our work the constant search time is achieved with local marking comparing to non-local extended marking of Coulomb potential.

In Sec.~\ref{sec:trapped} we analyze the time evolution of QRW states for a range of parameters in detail, and study the overlap of initial state with the eigenstates of the unitary operator corresponding to a single time step of QRW. Particularly, we demonstrate that the search speedup with constant time asymptotic is related to the appearance of trapped states of a specific structure near the defect: the two pairs of trapped states should simultaneously have a large matrix element with coordinate operator of the defect, and have large overlap with uniformly distributed initial state. In mathematical terms the trapped states are defined as a states that have compact support and zero density elsewhere. Their analogue in condensed matter are the strongly localized bound states on impurity in a system. We note that a number of mathematically rigorous studies have been made for spectral properties of quantum random walks and trapped states \cite{Inui2004PRA,Inui2005PRE,Endo2017,Fuda2021,Maeda2022,Fuda2023QS} together with their dynamics \cite{Fuda2017QIP,Mares2022PRA}. However, the possible efficiency of search algorithm based on exploiting the particular structure of QRW was not previously addressed.
The strong dependence of the trapped states structure on defect and walker parameters also suggests a scenario of protecting quantum search from a random disorder by an appropriate fine-tuning of marking parameters for the searched defect element. The corresponding study is presented in Sec.~\ref{sec:disorder} and the concluding remarks are given in Sec.~\ref{sec:conclusions}.
	
\section{The QRW model and localization phenomenon}
\label{sec:model}
 The quantum random walk analyzed in this work was first introduced in Ref.\cite{Kitagawa2010PRA} as a way to simulate topological insulators. Its single time step 
 \begin{align}
     \Psi(t+1)=U(\Theta_1,\Theta_2)\Psi(t)
 \end{align}
 is defined via three rotations of the spin and subsequent translations:
 \begin{align}\label{eq:QRW-definition}
 	&U(\Theta_1,\Theta_2)=T_{\vec{v}_3} R\left(\theta_1\right) T_{\vec{v}_2} R\left(\theta_2\right) T_{\vec{v}_1} R\left(\theta_1\right), \\ &R(\theta)=\left[\begin{array}{cc}
 		\cos (\theta / 2) & -\sin (\theta / 2) \\
 		\sin (\theta / 2) & \cos (\theta / 2)
 	\end{array}\right].\nonumber
 \end{align}
 The translation operations shift spin up components by $+\vec{v}_i$ and spin down by $-\vec{v}_i$. The vectors $\vec{v}_{1,2,3}$ can be defined on a triangular lattice as $v_{1}=(1,0)$, $v_{2}=(1/2,\sqrt{3/2})$ and $v_3=(1/2,-\sqrt{3/2})$, see inset in Fig.\ref{fig1-max-map}.

\begin{figure*}[t!]
\includegraphics[scale=1]{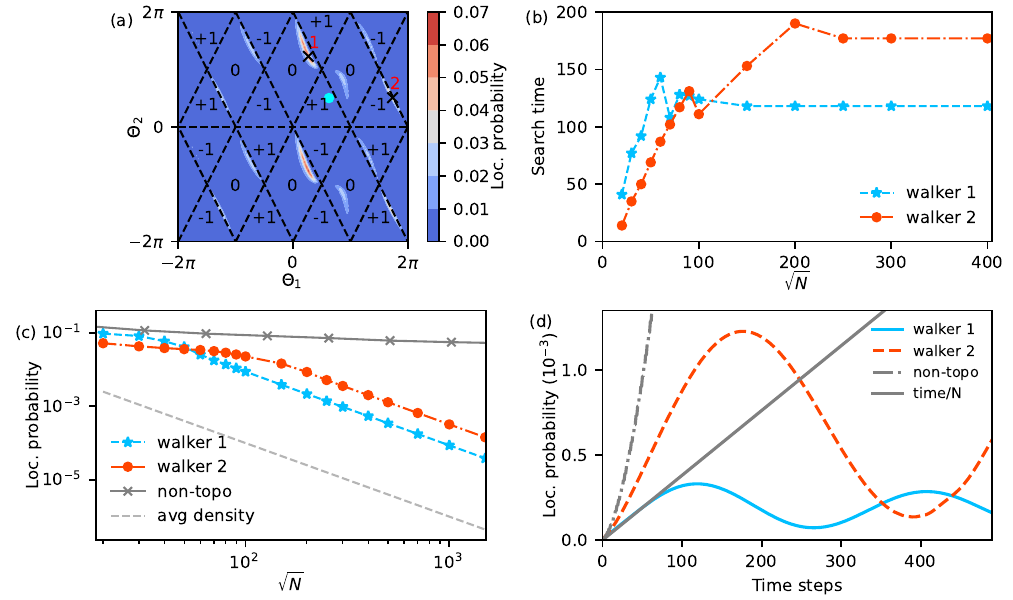}
        \caption{(a) A color map representation of a maximum localization probability on a defect site over the evolution time $T=1000$ for a system of size $N=40\times 40$ starting from a fully uniform distribution. Black numbers indicate Chern numbers of the corresponding parameter regions and dashed lines correspond to phase separation lines. Cyan dot marks the defect parameters $(5\pi/8, \pi/2)$. (b) Search time measured as a position of first probability peak as a function of system size for walker parameters indicated by crosses `1' and `2' in panel (a). (c) Localization probability dependence on system size. The two regions of different dependence are present where probability follows `Grover-type' scaling $P\sim 1/\log(N)$ as non-topological QRW from Ref.\cite{Roget2023PRR} until the time asymptotic changes to constant in (b) and then switches to system size scaling $P\sim 1/N$ but having few orders of magnitude larger value than average density in the system. (d) Comparison of time-dependent localization probability with non-topological QRW \cite{Roget2023PRR} and repeated classical search for system size $\sqrt{N}=512$. Walker $2$ shows advantage over classical. Note the scaling factor of $1e-3$ in the panel (d) for localization probabilities.}
        \label{fig2-colormaps-timeasyptotics}
    \end{figure*}

The spectrum of quasi-energies of this QRW without defects has two particle-hole symmetric bands due to the fact that matrix $U(\Theta_{1,2})$ is real. The corresponding states with energy $\pm E$ are related by complex conjugation $\Psi^{-E}=(\Psi^{E})^*$.
 Depending on parameters $\Theta_{1,2}$, such QRW realizes distinct topological phases with Chern numbers of Floquet-type bands $C=\pm 1,\, 0$. To make a contrast with the same kind of topological insulators \cite{Kitaev2009,Haldane1988} in crystalline solids, we note that QRW spectrum is defined as Floquet-type quasi-energies, and is thus limited to $(-\pi,\pi)$ interval. 
 The gap closing happens either at zero energy or at the ends of interval on the phase separation lines indicated in Fig.~\ref{fig2-colormaps-timeasyptotics}(a), and at $\Theta_2=0$ line gap closes everywhere \ref{app:properties-QRW-gaps}. 

In our setup the localization phenomena is expected to occur after certain amount of time steps at a single defect node on a finite 2D lattice of size $\sqrt{N}\times \sqrt{N}$ unit cells starting from initial state $|i\rangle$. We apply the periodic boundary condition in order to avoid the appearance of edge states. The  defect node is marked by parameters $\Theta_{1,2}^{def}$ that are different from surrounding $\Theta_{1,2}$. To estimate the possibility of given quantum walker to localize efficiently with high probability, we firstly scan the entire parameter range $\Theta_{1,2}$ for a given defect values $\Theta_{1,2}^{def}$. The measure of localization  probability that is relevant 
for quantum search applications is defined as a peak $\max_{t\in[0,T]}P_{def}(t)$ in time-dependent density of the system state at the defect node,
 \begin{align}
 P_{def}(t)=|\langle d|\Psi(t)\rangle|^2,
 \end{align}
 where $|d\rangle=\sum_{\sigma=\uparrow,\downarrow}|x=def,\sigma\rangle$ is the defect coordinate operator with both components of spin $\sigma$, and $|i\rangle$ is the initial state of the system defined as $|i\rangle=\frac{1}{\sqrt{2 N}}\sum_{\sigma,x}|x,\sigma\rangle$ with summation over entire system.
 The plot of the maximum value of density $P_{def}(t)$ on defect over the interval of time $T=1000$ for test system size  $N=40\times 40$ sites is shown in Fig.~\ref{fig1-max-map} for $\Theta_{1,2}^{def}=(5\pi/8,\pi/2)$. In Fig.~\ref{fig2-colormaps-timeasyptotics}(a) we show the color map representation of the same  localization probability distribution on the top of phase diagram in parameter space. For other positions of defect $\Theta_{1,2}^{def}$ in parameter space the colormaps of localization probability are shown in Supplemental material \cite{Supplement}. 
 
The islands of efficient  localization in parameter space fill a small areas placed in two distinct types of regions: inside topological phases and on phase separation lines. We compare the search time and localization probability behavior with growing system size for the walker parameters marked by `1' and `2' in the Fig.\ref{fig2-colormaps-timeasyptotics}(a), that realize  maximal values of probability within its regions. In Fig.~\ref{fig2-colormaps-timeasyptotics}(b), the search time is plotted against the square root of the system for the walkers '1' and '2'.
With the growing system size the behavior of search time demonstrates qualitative change - from $T\sim \sqrt{N}$ dependence to $T\sim \textit{const.}$ for both walkers. As we noted previously, such such a dependence of localization maximum on time was observed only for the Dirac QRW with particular type of Coulomb-like disorder potential \cite{Zylberman2021Entropy,Fredon2022Entropy}. In Fig.~\ref{fig2-colormaps-timeasyptotics}(c) the localization probability of walkers `1' and `2' is compared to average density at non-defect nodes. We find that the constant search time regime corresponds to the $P_{def}(t)= O(1/N)$ with a large numerical prefactor. For the algorithmic comparison it is important to note that this decrease of localization probability with system size $N$ is faster than 
 for non-topological QRW discussed in Ref.~\cite{Roget2023PRR}
$P_{def}^{Grover}=O(1/log\,N)$. The localization probability dependence in the form $O(1/N)$ arises due to the finite support of the trapped states. When the system size grows beyond the support of the trapped state, the search time asymptotic get saturated and we observe a clear dependence in the localization probability on the system size, see Fig.~\ref{fig2-colormaps-timeasyptotics}(c).
 
 For the fixed parameters of quantum walker `1' we visualized the distribution of localization probability in the space of defect parameters $\Theta_{1,2}^{def}\in [-2\pi, 2\pi]$. The corresponding discussion in Appendix~\ref{app:properties-QRW-gaps} shows that the localization probability depends only on $\Theta_1^{def}$ parameter, and is constant along the $\Theta_2^{def}$. Thus, it is more informative to present a dependence on $\Theta_{1}^{def}$ as shown in Fig.\ref{fig3:overlap-criteria}(a) to compare with overlap criteria introduced below.
 
 The asymptotic scaling of localization probability behaves as $O(const/N)$ in the large $N$ limit, where $\textit{const.}\in[100, 1000]$ is dependent on parameters of the quantum walker and defect. This asymptotic, in terms of algorithmic applications, is different from a fine-tuned regime in some non-topological QRWs, where probability could scale as $O(1/\log N)$ (see Ref.~\cite{Roget2023PRR}). At the same time, it outperforms the classical $1/N$ asymptotic by a constant factor. In Fig.\ref{fig2-colormaps-timeasyptotics}(d), we show how localization probability changes with time and compare to $\textit{time}/N$, which corresponds to classical search. Additionally, we compare to $1/\log N$ non-topological search. So while the topological QRWs do not reach quantum search Grover scaling, they can locally outperform classical search.
 In Sec.~\ref{sec:disorder}, we show that the topological QRW behavior shown in Fig.\ref{fig2-colormaps-timeasyptotics} can be further tuned by presence of disorder.

 \section{Contribution of trapped states to dynamical localization}
 \label{sec:trapped}
 To describe the origin of dramatically different localization behaviors for different parameters of the topological QRW  and to explain why the regions of efficient localization represent only a small fraction of the parameter space, we analyze the spectrum and localization properties of eigenvectors of a unitary operator of the system with one defect.

\begin{figure}
     \centering
\includegraphics[scale=1]{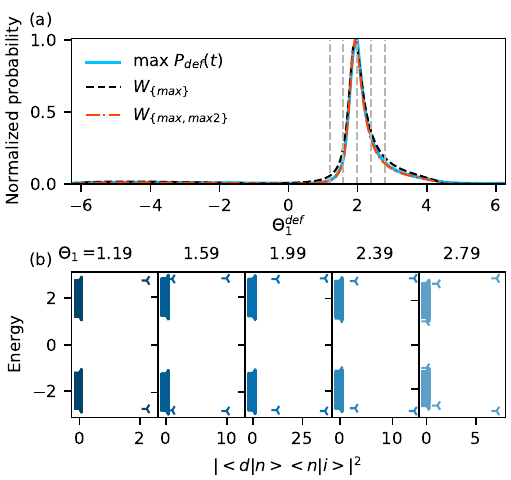}
     \caption{(a) Comparison of maximum value of probability on defect for a walker `1' normalized by numerically achieved maximum on the interval from Fig.\ref{fig2-colormaps-timeasyptotics} as a function of defect parameter $\Theta_1^{def}$ with the maximal value of squared sum of overlaps defined in Eq.\eqref{eq:definition-of-overlap-sum} with $\mathcal{M}$ containing one and two states respectively. All quantities are normalized to scale (0,1) for the purpose of comparison. The summation over two trapped states gives very accurate approximation of search probability peak.
     (b) The squared product of overlaps plotted as a function of energy of each eigenstate of operator $\bar{U}$ for walker parameters `1' and defect parameter $\Theta_1$ indicated on top of each panel. Trapped states differ from other states by significantly higher value of overlap.}
    \label{fig3:overlap-criteria}
 \end{figure}

 The probability evolution at a defect site at discrete times $t$ can be rewritten through the unitary operator of QRW,
 \begin{align}
     &P_{def}(t)
=|\langle d|\bar{U}^t(\Theta_1,\Theta_2)|i\rangle|^2 .
 \end{align}
The $\bar{U}$ notation is introduced in order to underline that unitary operator takes into account defect site. Substituting the eigenbasis decomposition of the $\bar{U}$ operator, $\bar{U}|n,\pm\rangle=e^{i \pm E_{n}}|n,\pm\rangle$, we have
\begin{align}\label{eq:P-def-via-eigenstates}
    P_{def}(t)=\bigg|\langle d|\sum_{n,\lambda=\pm}e^{i \lambda E_{n} t} |n,\lambda\rangle\langle n,\lambda|i\rangle\bigg|^2.
\end{align}
The saturation of search time to constant scaling after system size becomes large enough suggests that the properties of states that mainly contribute to the Eq.\eqref{eq:P-def-via-eigenstates} do not change with system size. This feature directly corresponds to the definition of trapped states previously introduced in Refs.~\cite{Inui2004PRA,Inui2005PRE} in the context of QRWs staying localized around its initial starting node over infinite evolution time. We investigate the contribution of different eigenstates of $\bar{U}$ to the evolution of probability, $P(t)$, in Fig.~\ref{fig3:overlap-criteria}(b) by comparing the product of overlaps  
$|\langle d|n,\lambda\rangle\langle n,\lambda|i\rangle|^2$. The panels in Fig.~\ref{fig3:overlap-criteria}(b) correspond to different parameters of defect (listed above each panel) denoted by vertical dashed lines for walker $`1'$ in Fig.~\ref{fig3:overlap-criteria}(a). 
The scale of $x$-axis in each panel of Fig.~\ref{fig3:overlap-criteria}(b) shows that defect parameters with higher localization probability also correspond to higher overlap product for several states - in this case two pairs of particle-hole symmetric states. To further verify this observation, we compare the maximal value of overlap product 
\begin{align}
   \label{eq:definition-of-overlap-sum} W_{\mathcal{M}}=\left|\sum_{n\in \mathcal{M},\lambda=\p}\langle d|n,\lambda\rangle\langle n,\lambda|i\rangle\right|^2
    \end{align}
    of a single eigenstate of $\bar{U}$ with the normalized probability dependence $max[P_{def}(t)]$ as function of $\Theta_{1}^{def}$. The comparison is made by taking a single pair of states that maximize overlap with set of indices $\mathcal{M}=\{n_{max}\}$, as well as two pairs of states $\mathcal{M}=\{n_{max}, n_{second\,max}\}$. The results presented in Fig.~\ref{fig3:overlap-criteria}(a) indicate that while $\mathcal{M}=\{n_{max}\}$ selection overestimates the region of good search, the $\mathcal{M}=\{n_{max}, n_{second\,max}\}$ very precisely describes the high search probability peak. To further check this correspondence between the evolution of described by two pairs of states with maximal overlap and the maximum probability distribution, we show the same comparison for a defect $\Theta_{1,2}^{def}=(5\pi/8,\pi/2)$ along phase separation line and the line with walker $`1'$ in  Appendix \ref{app:comparison-walker-2}. The correspondence for a test system size with $\sqrt{N}=40$ shows that efficient search for such split-step QRW is mostly determined by the appearance of two pairs of trapped states that have large overlap with both defect and initial state.

Finally, we perform a comparison of time evolution defined by two pairs of trapped states that maximize overlap criteria with the evolution of probability on the defect. For this purpose we note that the $\bar{U}$ has the same particle-hole symmetry as $U(\Theta_{1,2})$ because it is a real matrix. In addition, by calculating overlap products denoted as $w_{\pm, j}^{\sigma}=\langle d,\sigma|j\rangle\langle j|i\rangle$ for each spin $\sigma=\uparrow,\downarrow$ and pair of trapped states $j$ at the defect site separately, we find that for each trapped state there is a phase difference $w_{+, j}^{\downarrow}=iw_{+, j}^{\uparrow}$ and corresponding $w_{-, j}^{\downarrow}=-iw_{-, j}^{\uparrow}$ up to numerical precision. The former symmetry is present only at defect site and is a property of trapped states localized around the defect. Thus, evolution of both spin components at the defect site that is generated by trapped states is described as follows
\begin{align}
\label{eq:spin-up-component-on-defect-psi}
&\Psi_{\uparrow,d}(t)=2\sum_{j=1,2}\left|w_{+, j}^{\uparrow}\right| \cos \left(E_j t+\arg \left(w_{+, j}^{\uparrow}\right)\right),\\
\label{eq:spin-down-component-on-defect-psi}
&\Psi_{\downarrow,d}(t)=2\sum_{j=1,2}\left|w_{+, j}^{\uparrow}\right| \sin \left(E_j t+\arg \left(w_{+, j}^{\uparrow}\right)\right),
\end{align}
where we took into account the particle-hole symmetry relation $w_{-,j}^{\sigma}=(w_{+,j}^{\sigma})*$. It is important to emphasize that after calculating probability at a defect site, one finds that the two spin components exactly cancel oscillating parts of each other for a single value of $j$. That happens due to the different phases of oscillations between $sin$ and $cos$ in Eqs.\eqref{eq:spin-up-component-on-defect-psi} and \eqref{eq:spin-down-component-on-defect-psi} above, and results in:
\begin{align}
    &|\Psi_{\sigma,d}^{j}(t)|^2 = 2\left|w_{+, j}^{\uparrow}\right|^2\nonumber\\
    &\times\left(1+\text{sign}(\sigma) \cos \left(2 E_j t+2 \arg (w_{+, j}^{\uparrow})\right)\right),
\end{align}
with $\text{sign}(\sigma)=\pm$. Thus, the contribution of each single pair of trapped states is constant and due to the initial condition of uniformly distributed state over entire system it could not describe the high peak in localization probability. 
 \begin{figure}
     \includegraphics[scale=1]{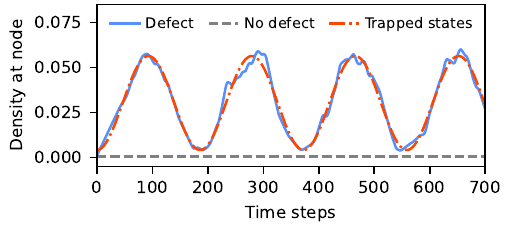}
     \caption{The comparison of density evolution at searched defect node (blue solid line) and another distanced node without defect (gray dashed line) with time for a walker parameters `1' and defect marked by $(5\pi/8, \pi/2)$. The red dash-dotted line corresponds to the evolution described by only two pairs of trapped states that maximize product of overlaps and given by Eq.~\eqref{eq:evolution-via-trapped-states-full}. The evolution from trapped states approximates with very good precision the first few peaks in probability and the time of their occurrence.}
     \label{fig4:evolution-vs-trapped-states}
 \end{figure}
 However, the crucial contribution comes from the superposition between two pairs of trapped states and is manifested in the last oscillating term in the total density evolution at the defect: 
\begin{align}
\label{eq:evolution-via-trapped-states-full}
&\left|\Psi_{\uparrow,d}(t)\right|^2+\left|\Psi_{\downarrow,d}(t)\right|^2=4\left(\left|w_{+, 1}^{\uparrow}\right|^2+\left|w_{+, 2}^{\uparrow}\right|^2\right)+8\nonumber\\
& \times\left|w_{+, 1}^{\uparrow}\right| \left|w_{+, 2}^{\uparrow}\right|\cos \left(\left(E_1-E_2\right) t+\arg w_{+, 1}^{\uparrow}-\arg w_{+, 2}^{\uparrow}\right).
\end{align}
The period of this oscillation is different from the fast oscillations of each pair of trapped states and is defined by the energy difference between the positive energy levels of two pairs of trapped states $(E_{1}-E_{2})$.  
Fig.~\ref{fig4:evolution-vs-trapped-states} shows that expression in Eq.~\eqref{eq:evolution-via-trapped-states-full} captures very well the structure of oscillations of localization probability despite its simplicity.
In Fig.~\ref{fig4:evolution-vs-trapped-states} the results are presented for a walker `1'. We show the equivalent result walker `2' in Appendix \ref{app:comparison-walker-2}. To further verify the agreement of predicted period of oscillations with the actual oscillation of probability on a defect node we used numerical Fourier transform. The agreement with period defined by $(E_{1}-E_{2})$ has errors below $2 \%$ on the timescale of evolution $T=5000$ for system with $\sqrt{N}=40$. 
It was observed \cite{Wong2018,Roget2023PRR} that in non-topological QRWs, that realize localization mechanism similar to Grover's algorithm, and, the period of oscillation is defined by an energy difference within the same particle-hole symmetric pair of states at lowest energies. The difference between two definitions of the period shows that the mechanism of localization in topological QRWs is different comparing to non-topological QRWs studied in literature.

The change of time asymptotic that is shown in Fig.~\ref{fig2-colormaps-timeasyptotics}(b) happens when the system size becomes larger that diameters of both trapped state pairs. In  Appendix \ref{app:structure-of-trapped-states}
we show that the state from the pair with smaller overlap in Fig.\ref{fig2-colormaps-timeasyptotics}(c) has larger radius of localization ($95\%$ of the probability density). This pair of states determines the chance in the search time asymptotic as shown in Fig.~\ref{fig2-colormaps-timeasyptotics}b.

\begin{figure*}
    \includegraphics[scale=1]{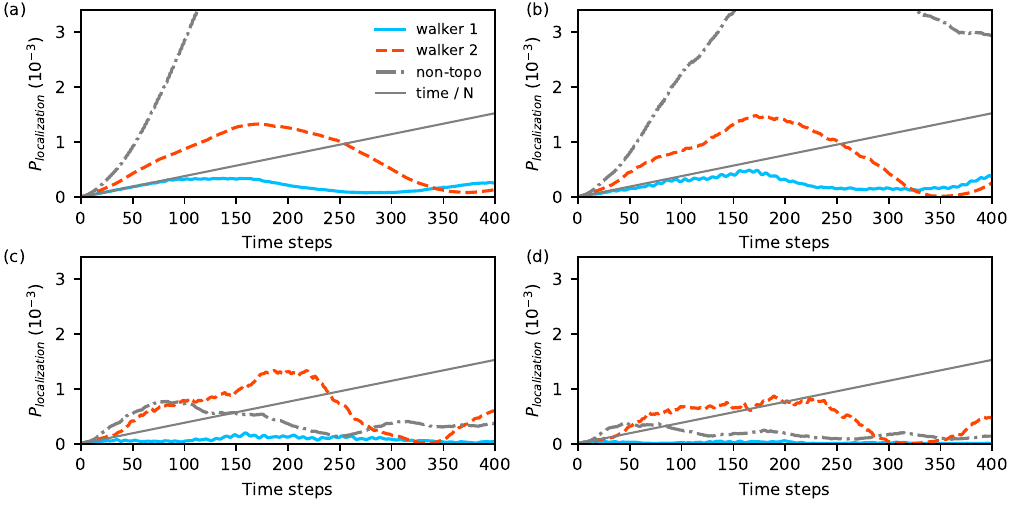}
    \caption{The localization probability calculated for 2 walker parameters analyzed in the main text (see Fig.2) and compared with the localization probability for QRW from Ref.\cite{Roget2023PRR} and with classical search with linearly increasing probability. The system size is $512\times 512$. Disorder strength varies from small in panel (a) $\Theta_{dis}=0.05$, to (b) $\Theta_{dis}=0.15$, (c) $\Theta_{dis}=0.35$, and large in panel (d) $\Theta_{dis}=0.5$. The walker $2$ shows best performance at large disorder out of all compared search procedures.}
    \label{figS:disorder-protection}
\end{figure*}

\section{Dynamical localization on defect in the presence of disorder}
\label{sec:disorder}

\begin{figure}
    \includegraphics[scale=1]{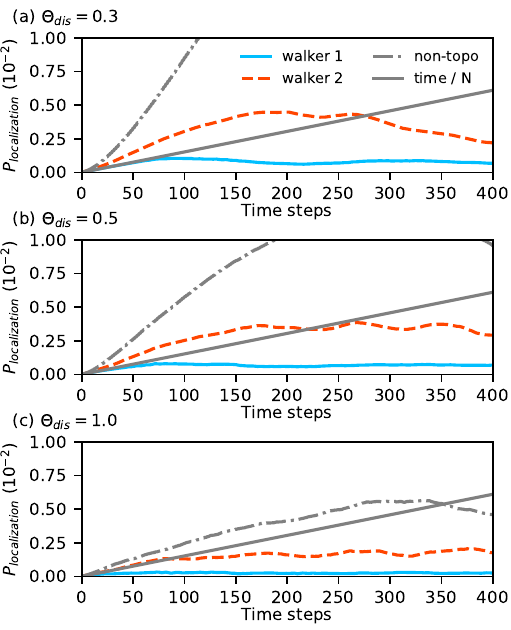}
    \caption{The dynamical localization phenomena on a single-node defect averaged over $20$ configurations of disorder. The averaged localization in non-topological split-step QRW from Ref.\cite{Roget2023PRR} has a higher density of defect node rather than of a topological QRW due to a larger amount of states contributing to maximum localization. The size of the system is $256\times 256$ nodes.}
    \label{fig:disorder-averaged}
\end{figure}

Now we move on to analyzing the stability of localization on single-node defect with disorder. For the topological QRW the disorder is added to rotation operators on the nodes in Eq.\eqref{eq:QRW-definition}, via replacement $R(\Theta)\to R(\Theta+\delta\Theta)$ where 
$\delta \Theta \in [-\Theta_{dis}, \Theta_{dis}]$ is a uniformly distributed random value centered around zero in finite interval. The defect node is not disordered. As an example of another mechanism, the  localization of non-topological discrete-time QRW from Ref.~\cite{Roget2023PRR} is studied for the same system sizes. 
The corresponding time step operator is defined as:
\begin{align}
    U=T_y C_y T_x C_x,\quad T_{i}|\pm\rangle |i\rangle= |\pm\rangle |i\pm 1\rangle.
\end{align}
In this case the disorder is implemented as phases in the coin rotation operators: 
\begin{align}
    &C_x(\delta\Theta)= \frac{1}{\sqrt{2}}\begin{pmatrix}
        1 & i \exp(i \delta \Theta) \\
        i \exp(-i \delta \Theta) & 1
    \end{pmatrix}, \\ 
    & C_y(\delta\Theta)= \frac{1}{\sqrt{2}}\begin{pmatrix}
        1 & -i \exp(i \delta \Theta) \\
        -i \exp(-i \delta \Theta) & 1
    \end{pmatrix}. 
\end{align}
While different selections of disorder could be made, the one introduced above is motivated by the possible loss of phase coherence on the nodes of large system (described by slightly different phases acquired by passing each node).
 The sizes of the test systems are taken to be the same - $512\times 512$ nodes for single configuration of disorder, and $256\times 256$ for averaging over $20$ configurations. Both sizes are well above the saturation time threshold for example walker parameters `1' and `2' described in the previous sections. 

 The results of simulation for different sizes of interval distributions $\delta \Theta \in [-\Theta_{dis}, \Theta_{dis}]$ of angle disorder and for single configuration in coordinate space are presented in Fig.~\ref{figS:disorder-protection}. The values $\Theta_{dis}$ are equal for topological and non-topological QRWs. 
The disorder reduces the height of localization probability peak for both quantum walks, and consequently the corresponding efficiency of search algorithms, as it is shown in panels (a)-(d) for $\Theta_{dis}$ varying from $0.05$ to $0.5$.. 
However, for one of the two topological walkers, the search probability always remains higher that the classical search probability during the first period of oscillation. Notably, the probability of quantum search (non-topological QRW) drops dramatically as the disorder increases.

While the example shown in Fig.~\ref{figS:disorder-protection} constitutes one specific disorder configuration and type, the trapped states demonstrate relatively stable behavior, even locally outperforming quantum search for large disorder. Still, we point out that the more systematic study should be made to define the optimal performance conditions of different quantum  and classical search algorithms.
One should note, however, that our algorithm still outperforms classical search in localization probability maximum at fixed time. The classical search always behaves as $P_{classical}(t)=t/N$ with $t$ - number of check ups of individual nodes.

Finally, the three panels of Fig.\ref{fig:disorder-averaged} demonstrate a study of disorder effect on two mechanisms of localization with average over 20 spatial configurations. We show three different values of disorder interval limit $\Theta_{dis}\in\{0.3,0.5,1.0\}$. The small oscillations of the localization probability are not present in these plots, and the curves are more smooth. The  effect of reducing the maximum value of $P_{localization}$ is present, especially for large disorder with $\Theta_{dis}=1$ in panel (c). The localization of non-topological QRW demonstrates a higher probability than both example walkers from topological QRW.

\section{Conclusions}
\label{sec:conclusions}
In conclusion, we demonstrated that a fast dynamical localization phenomenon takes place in the split-step discrete-time topological quantum random walks. It was shown that localization mechanism is based on just a two pairs of localized trapped states associated with defect. Each pair of trapped states is particle-hole symmetry in the quasienergy spectrum. The time of highest localization is defined by the gap that separates two positive-energy trapped states. Due to a finite radius of localization, the position of these states in spectrum is not influenced in large enough systems. We should note that the overlap product criteria served as efficient tool to identify these trapped states, capture the evolution of walker and uncover the principle behind distribution of high-localization peaks in parameter space of walker.

The described phenomena open an avenue towards realization of the search algorithms based on topological QRWs as well as further studies on explainability of localization phenomena in quantum random walks. The search mechanism, as formulated in this work, can be immediately tested in photonic and synthetic lattice quantum walk experiments that realize topological QRW \cite{Kitagawa2012NatComm,Chen2018PRL,Nitsche2019NJP,Esposito2022NPJ}.

\begin{acknowledgements}
We benefited from discussions with Tereza Vakhtel, Tomas Osterholt, Anton Akhmerov, Yaroslav Herasymenko, Margarita Davydova and Carlo Beenakker.
This project was supported by the Kavli Foundation. This publication is part of the project Engineered Topological Quantum Networks (with Project No. VI.Veni.212.278) of the research program NWO Talent Programme Veni Science domain 2021 which is financed by the Dutch Research Council (NWO). GJ acknowledges the research program “Materials for the Quantum Age” (QuMat) for financial support. This program (registration number 024.005.006) is part of the Gravitation program financed by the Dutch Ministry of Education, Culture and Science (OCW).
\end{acknowledgements}

\begin{center}
{\bf Code and data}
\end{center}
The code to reproduce research in this paper can be found at this URL: \href{https://gitlab.com/QMAI/papers/topologicalsearch}{https://gitlab.com/QMAI/papers/topologicalsearch}. The Zenodo repository containing both code and data can be found at this URL: \href{https://zenodo.org/records/14853734}{https://zenodo.org/records/14853734}.



\newpage
\appendix
\section{Properties of split-step quantum random walk}
\label{app:properties-QRW-gaps}
\begin{figure}
    \centering
\includegraphics[scale=0.55]{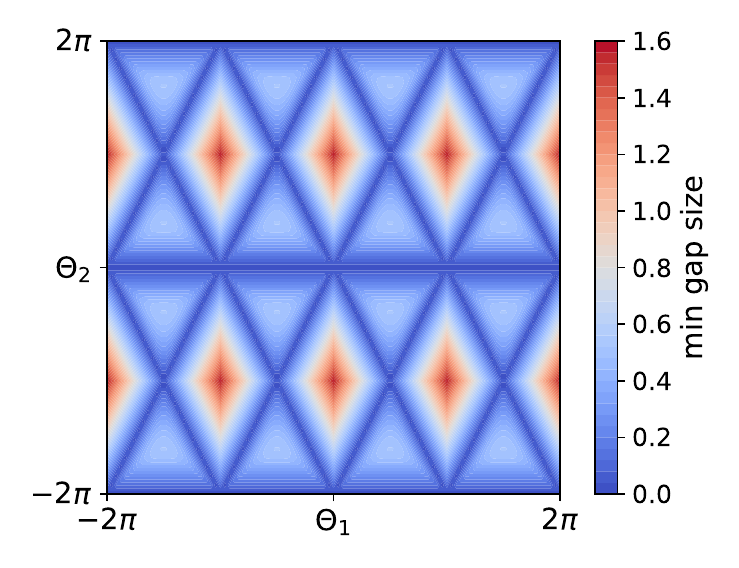}\\
\includegraphics[scale=0.55]{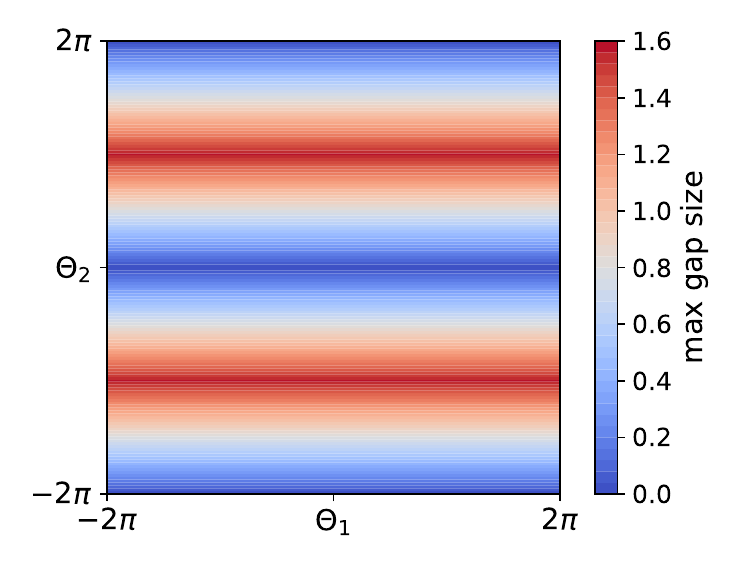}
    \caption{The regions of one gap closing (upper panel) and both gaps closing (lower panel). Note the agreement between the upper panel and phase separation lines position in Fig.~\ref{fig2-colormaps-timeasyptotics}a.}
    \label{fig-s:gap-closing-regions}
\end{figure}

In this section we present a number of properties for split-step QRW that realizes topological phases in the case without defect nodes. Using Eq.(1) from the main text, and assuming the infinite system in both directions, we find the following equation for spectrum of translation-invariant unit cell:
\begin{align}
	&\lambda ^2+2 \lambda  \cos \frac{\Theta _2}{2} \left(\sin ^2\frac{\Theta_1}{2} \cos \left(2 k_y\right)-\cos ^2\frac{\Theta _1}{2} \cos
	\left(k_x\right)\right)\nn
    &+\lambda  \sin \left( \Theta _1\right) \sin \frac{\Theta
	_2}{2} \left(\cos \left(k_x-2 k_y\right)+1\right)+1=0.
\end{align}
The solution of this equation, $\lambda_{ \pm}=e^{ i E}$, gives a particle-hole symmetric spectrum of quasi-energies. The quasi-energy is given by
\begin{align}
	&E=\pm \arccos \frac{b}{\sqrt{2b^2-4}},\nn
	&b=  2  \cos \frac{\Theta _2}{2} \left(\sin ^2\frac{\Theta
	_1}{2} \cos \left(2 k_y\right)-\cos ^2\frac{\Theta_1}{2} \cos
	\left(k_x\right)\right)\nn
    &+  \sin  \Theta _1 \sin \frac{\Theta
	_2}{2} \left(\cos \left(k_x-2 k_y\right)+1\right)
\end{align}
The topological phases and corresponding Chern numbers were defined in Ref.\cite{Kitagawa2010PRA}. The phase separation lines are characterized by the gap closing. For the technical purposed of identifying different types of search speed up discussed in the main text, we plot a map of gap value: the minimal gap value between two bands at $E=0$ or $E=\pm\pi$ (see Fig.\ref{fig-s:gap-closing-regions}(a) ) and the maximal gap value around the same energies. From such maps we find that the full gap closing happens only at lines $\Theta_{2}=2\pi n$ with $n$ being integer.

\begin{figure}
    \centering
    \includegraphics[scale=0.55]{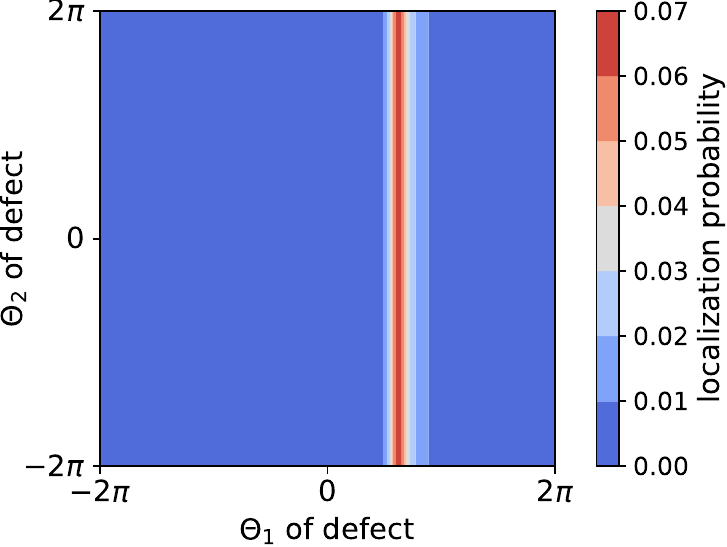}
    \caption{2D color map of dependence on defect parameters for a walker `1'. The optimal search is performed on a line - which demonstrates that the dependence on $\Theta_{2}^{def}$ is absent.}
    \label{figs:defect-dependence}
\end{figure}

The colormap of search dependence on defect parameters for the walker `1' shown in Fig.2 of the main text is presented in Fig.\ref{figs:defect-dependence}. The dependence on $\Theta_{2}^{def}$ is absent, which motivates the studying of only different $\Theta_1^{def}$ values in the main text.

\section{The comparison of product of overlaps with optimal localization regions}
\label{app:comparison-walker-2}
\begin{figure}
    \centering
    \includegraphics[scale=0.48]{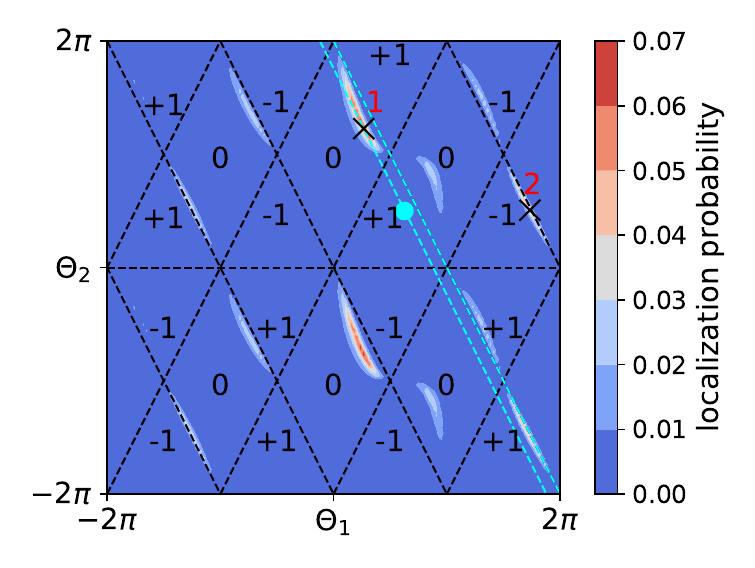}
    \includegraphics[scale=0.43]{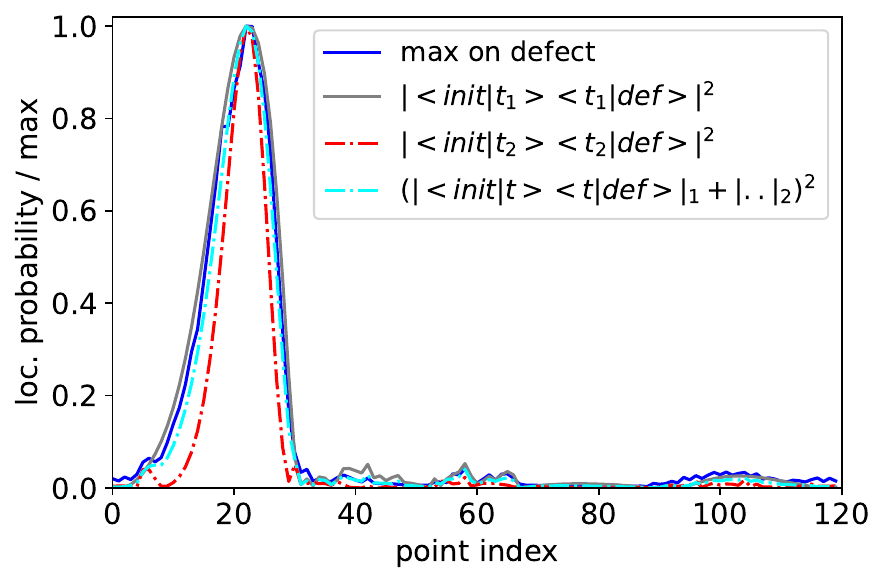}
    \includegraphics[scale=0.43]{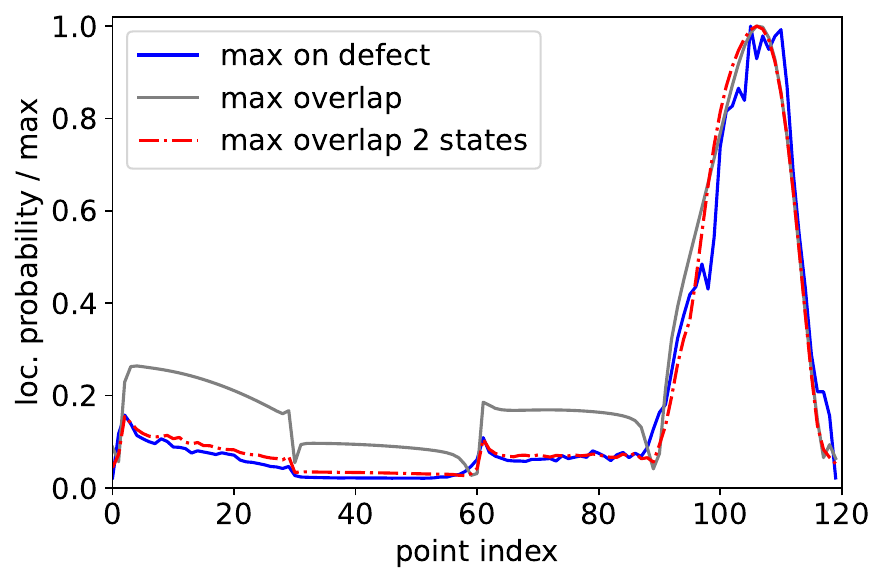}
    \caption{Comparison of maximized over time localization probability with maximal product of overlaps described in the main text found over all eigenstates of evolution operator $\bar{U}$ for a two lines in parameter space plotted as dashed cyan lines in the first panel. The second panel corresponds to a line going through walker $`1'$ and the third panel corresponds to a phase separation line. }
    \label{figS:overlap-criteria-lines}
\end{figure}

In this Appendix we analyze how precise the correspondence between maximization of $|<d|n><n|i>|^2$ and $(\sum_{n=max_1, max_2}|<d|n><n|i>| )^2$ is compared to the regions of optimal localization in parameter space for a fixed defect parameters. For this purpose we compare the max search probability curves plotted along two lines depicted as dashed cyan in the first panel of Fig.\ref{figS:overlap-criteria-lines}. One of these lines is a phase separation line and second one is parallel to phase separation and contains a walker $`1'$ discussed in the main text. This second line crosses different topological phases. 
In the second and third panels of Fig.\ref{figS:overlap-criteria-lines} we plot the comparison of max probability found on time interval $T=1000$ for the system size with $\sqrt{N}=40$ with values of different overlaps. The second and third panels in Fig.\ref{figS:overlap-criteria-lines} clearly demonstrate that the combination of maximum and second maximum products of overlaps - the contribution of two pairs of trapped states - describes the probability peak most precisely. This leads to a conclusion that efficient search is performed by two pairs of trapped states in this system in all cases studied. 

\begin{figure}
    \centering
    \includegraphics[scale=0.38]{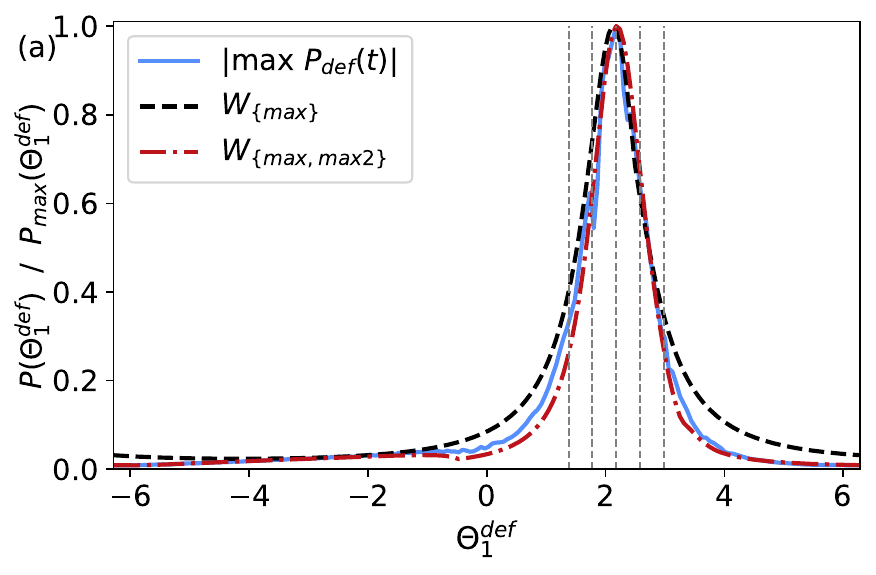}
    \includegraphics[scale=0.4]{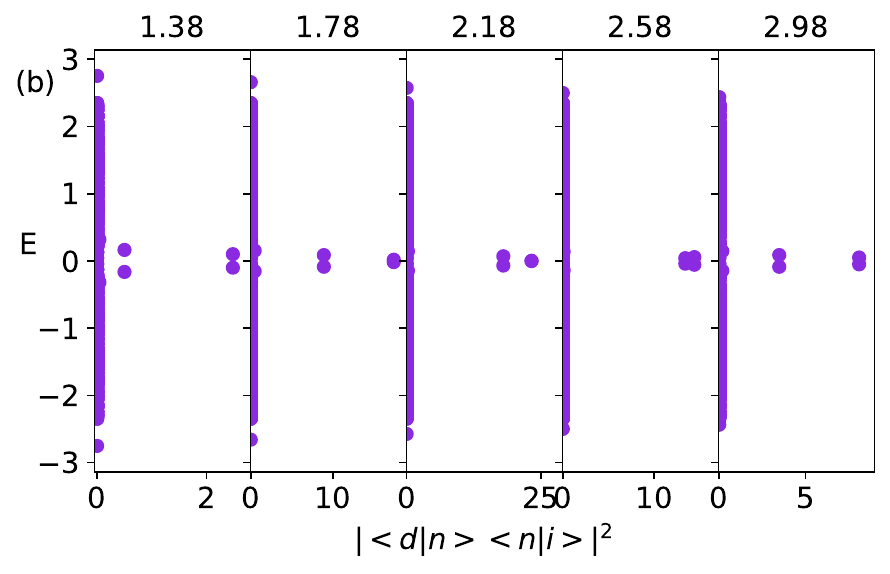}
    \includegraphics[scale=0.38]{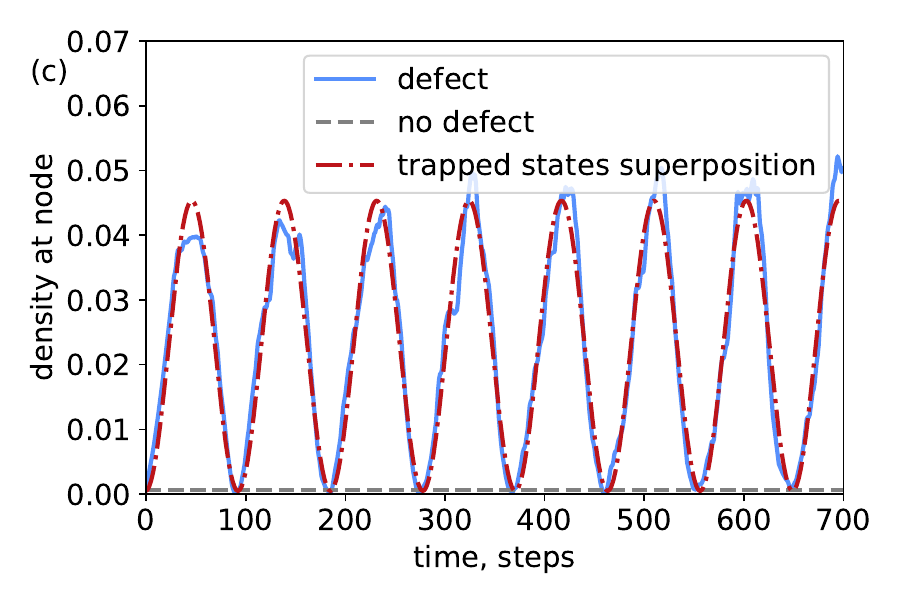}
    \caption{(a) The comparison of maximized over time localization probability with overlap criteria for walker `2', with all quantities being normalized by their maximum value on the same grid of $\Theta_1^{def}$ parameters. (b) The identification of states that mainly contribute into constructive superposition on a searched defect node for walker `2' parameters. (c) The comparison of density evolution at defect node (blue solid line) with the superposition (red dashed line) of trapped states identified from panel (b).}
    \label{figS:evolution-for-second-walker}
\end{figure}

Next, we analyze the structure of trapped states and their position in spectrum for the example walker parameters labeled as $`2'$. The analysis is similar to what is done in the main text and is presented in three panels of Fig.\ref{figS:evolution-for-second-walker}: comparison of search probability with overlap criteria in the first panel, the set of overlaps for each eigenstate of $\bar{U}$ in second panel and the comparison of probability evolution at the defect with the one defined by the two pairs of the trapped states in the third panel. However, in this case we find that the two pairs of trapped states that mainly contribute to the efficient search are placed near zero energy. Also there is no gap in spectrum around this energy, but the in-gap trapped states placed around $E=\pm \pi$ do not contribute into the constructive interference relevant for successful search. 

\section{Structure of trapped states in coordinate space}
\label{app:structure-of-trapped-states}
\begin{figure}
    \centering
\includegraphics[scale=0.9]{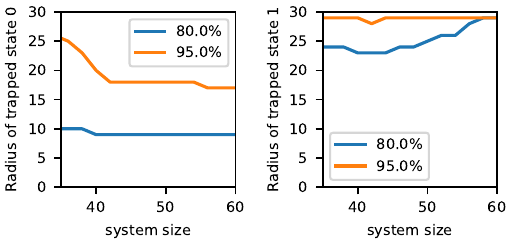}
    \caption{Radius of trapped state as a function of system size. Left panel corresponds to a trapped state with largest overlap $|\langle d|n\rangle\langle n|i\rangle|$ in Fig.3(b) of the main text, and right panel corresponds to the state with second largest overlap. Two curves correspond to the radius of area in which $80\%$ and $95\%$ of state density concentrated.}
    \label{figS:radius-dependence}
\end{figure}

In this section we analyze how extended are the trapped states that mainly contribute to the search algorithm. From the search time dependence on system size shown in Fig.2(b) in the main text, we conclude that the more extended trapped state has radius $R\approx 50$ for walker `1' and $R\approx 100$ for walker `2' and defect parameter $\Theta_{1}^{def}=5\pi/8.$. The radius of trapped state was estimated as twice smaller that the system size where transition in time asymptotic happens. 
To further analyze the trapped state radius dependence on system size, we plot it in Fig.\ref{figS:radius-dependence}. The set of tested system sizes covers the region where the trapped states are having overlap with itself due to periodic boundary condition. To reduce the numerical precision issue related to identifying density of quickly decaying state far from its center, we plot a dependence of radius which covers $80\%$ and $95\%$ of density. Two plots show results for a state taken from each pair of trapped states discussed in the main text in Fig.3(b).  The states from second pair have larger radius and it stronger depends on system size. Thus, the change in search time asymptotic is mostly defined by this second pair of trapped states, that has second maximal overlap value.

\newpage
 \bibliography{grover_bib}


\end{document}